\documentstyle[11pt,newpasp,twoside,epsf]{article}
\def\edcomment#1{\iffalse\marginpar{\raggedright\sl#1\/}\else\relax\fi}
\reversemarginpar

\begin{document}
\title{Evolution  of galaxies in triaxial halos with figure rotation }
\author{Kenji Bekki}
\affil{School of physics, University of New South Wales, Sydney, NSW, 2052, Australia
}
\author{K. C. Freeman}
\affil{Mount Stromlo and Siding Spring Observatories, The Australian National University,
Private Bag, P.O. Weston Creek, ACT 2611, Australia}

\begin{abstract}

 Firstly, we demonstrate that  unusually large outer HI spiral arms
observed in NGC 2915 can  form 
in an extended  gas disk embedded in a massive triaxial dark
matter halo with slow figure rotation, through the strong gravitational
torque of the rotating halo.  
Secondly, we  
show that the figure rotation
of a triaxial dark matter halo can influence dynamical evolution of disk galaxies
by using fully self-consistent numerical simulations.
We particularly describe the formation processes of 
``halo-triggered'' bars in thin galactic disks
dominated by dark matter halos with figure rotation 
and discuss the origin of stellar bars in low luminosity, low
surface brightness (LSB) disk galaxies. 
Thirdly, we provide several implications of the present numerical
results in terms of triggering mechanism of starbursts in galaxies and stellar bar
formation in high redshifts. 

\end{abstract}

\section{Kinematics of dark matter halos and galaxy evolution} 

Although several attempts have been so far made to reveal the {\it shapes} 
(e.g., the degree of oblateness or triaxiality) of dark matter halos
in galaxies 
(e.g., Sackett \&  Sparke 1990; Franx, van Gorkom, \& de Zeeuw 1994;
Olling 1995; Salucci \& Persic 1997; Sackett 2003 in this conference),
{\it rotational properties} of dark matter halos
have been less discussed.  Based on the detailed
analysis of structure and kinematics of the very extended HI disk
around NGC 2915, Bureau et al (1999) first suggested that the observed
spiral-like structures in the HI disk can be formed  by a triaxial halo
with figure rotation. 
However, it is unclear whether a triaxial dark
halo with figure rotation is really responsible for the observed
extended spiral structures in NGC 2915, because of the lack of
numerical studies of gas dynamics in the gravitational potentials of
triaxial halos with figure rotation.
Also it is an important problem to understand {\it how figure rotation
of a dark matter halo in a disk galaxy influences dynamical evolution,
star formation history, and chemical evolution of the disk},
because previous theoretical/numerical studies  so far 
have not extensively investigated such influence of figure rotation
of dark matter halos. Therefore, we here investigate numerically
(1) whether the observed NGC 2915's giant HI spirals  
can be formed by tidal force of the figure rotation of the triaxial dark halo
and (2) how the figure rotation of a triaxial dark halo in a disk galaxy
influences the formation of stellar bars and spiral arms within it.

\section{Evidences of a triaxial halo with figure rotation ?} 

We consider an extended uniform  gas disk  of a dwarf galaxy
embedded by a triaxial dark matter halo with figure rotation. 
We adopt TREESPH codes described in  Bekki (1997) for hydrodynamical evolution
of galaxies and thereby investigate the dynamical evolution  of the gas disk
under the triaxial dark halo (The details of the model are given
in Bekki \& Freeman 2002).
The unusually extended HI gas in NGC 2915 
is modeled as an uniform thin gas disk with
the size (represented by $R_{\rm g}$) of 15 kpc and the mass ($M_{\rm g}$) of 10$^8$ $M_{\odot}$.
The gas disk with uniform radial density distribution is represented by
20K SPH particles and each gas particle is 
first placed in the $x$-$y$ plane and given its circular velocity 
(determined by the dark matter halo) at its radius.
The gas disk is then inclined by ${\theta}$ degree  with respect to the $x$-$y$ plane. 
An  isothermal equation of state is used for the gas
with a temperature of $1.2\times 10^3$ K corresponding to a sound speed
of 4 km $\rm s^{-1}$ ($\sim$ 0.075 times the virial velocity of the system).
 This extended and inclined gas disk is assumed to be dynamically affected
 {\it only} by a massive dark matter halo with the mass $M_{\rm DM}$ of
 10$^{10}$ $M_{\odot}$ and
 we adopt the density distribution of the NFW 
 halo (Navarro, Frenk \& White 1996) suggested from CDM simulations:
 \begin{equation}
 {\rho}(r)=\frac{\rho_{0}}{(r/r_{\rm s})(1+r/r_{\rm s})^2},
 \end{equation} 
 where  $\rho_{0}$ and $r_{\rm s}$ are the central density and the scale
 length of a dark halo, respectively.  
 We take the isodensity
 surfaces of the dark halo to be triaxial ellipsoids on which the
 Cartesian coordinates ($x$,$y$,$z$) satisfy the following condition
 (Binney \& Tremaine 1987):  \begin{equation} m^2 \equiv
 \frac{x^2}{a^2}+\frac{y^2}{b^2}+\frac{z^2}{c^2} = constant,
 \end{equation} where  $a$, $b$, and $c$ are the parameters which
 determine the two axis ratios of a triaxial body (i.e., long to short
 and long to middle).  In the present study, $a$ is set to be 1 and the
 long-axis is initially coincident with the $x$ axis. Accordingly $b$
 ($\le $ 1) and $c$ ($\le $ 1) are free parameters which determine the
 shapes of triaxial dark matter halos.  The triaxial halo is assumed to
 be rotating as a solid body with the pattern speed of ${\Omega}_{\rm
 p}$.  
By changing the parameters $R_{\rm g}$, $\theta$, $b$, $c$, and ${\Omega}_{\rm p}$,
we investigate morphological evolution of extended gas disks and its dependence
on shapes and rotational properties of triaxial halos.
The parameter values of $R_{\rm g}$, $\theta$, $b$, $c$, and ${\Omega}_{\rm p}$
are 15 kpc, 30$^{\circ}$,  0.8, 0.6,  
and  3.84 km s$^{-1}$ kpc$^{-1}$,
respectively, in  {\it the standard model} (Model 1).
In the following, our units of mass, length, and time are
$10^{10}$ M$_{\odot}$ (= M$_{\rm DM}$),
15 kpc, 
and 2.74 $\times$ $10^8$ yr, respectively.

\begin{figure}
\plotone{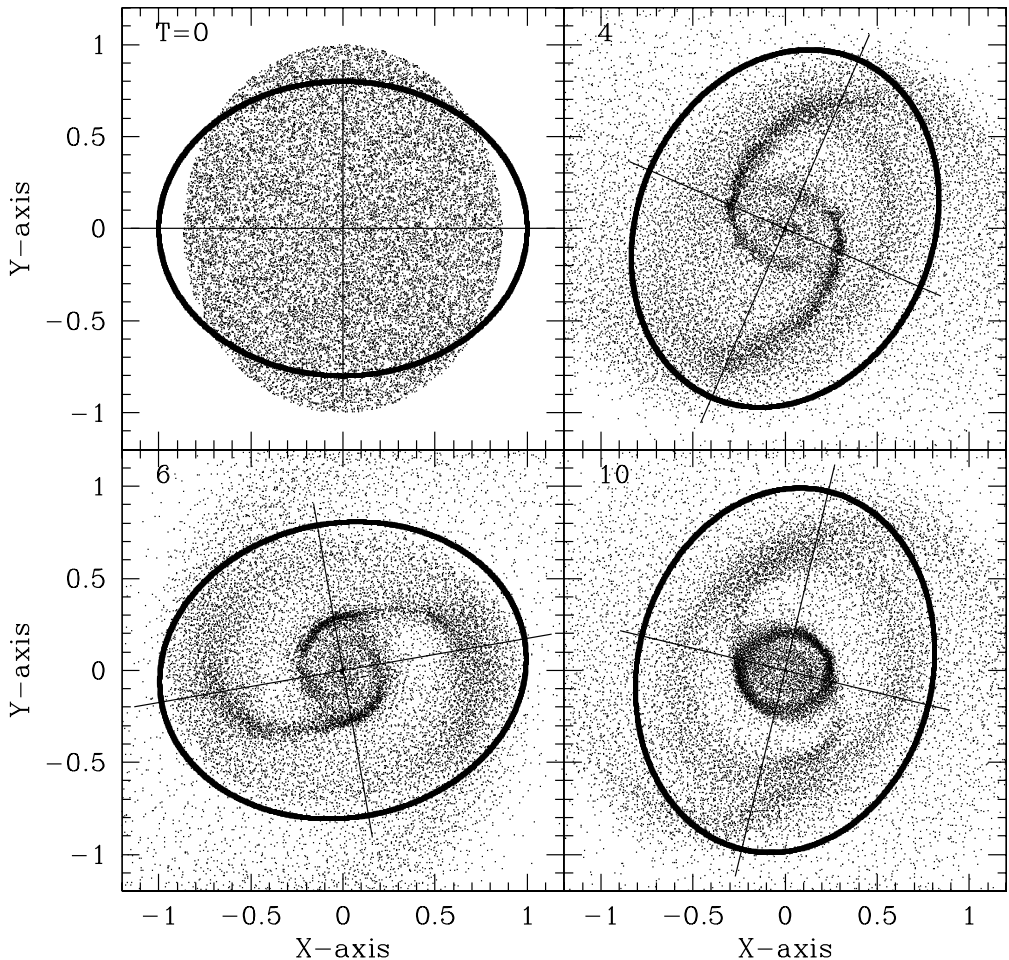}
\caption{
Morphological evolution of the gas disk projected onto the $x$-$y$ plane
for the standard model with the disk inclination ($\theta$) of 30$^{\circ}$
with respect to the $x$-$y$ plane. 
The time ($T$) indicated in the upper left
corner of each frame is given in our units (2.74 $\times$ $10^8$ yr) and
each frame (2.4 in our units) measures 36 kpc on a side.
The shape of the triaxial halo projected onto the $x$-$y$ plane (at $R$ = 1.0 in our units,
corresponding to 15 kpc)
is outlined by a thick solid line at each time $T$. 
Both the gas disk and the halo are assumed to rotate counter-clockwise
(The spin axis of the halo is coincident with the $z$ axis).
The long and middle axes  of the halo are represented by thin solid lines.
Note that as the triaxial halo rigidly rotates, two open trailing spiral arms
are formed at $T$ = 4.0 ($\sim$ 1.1 Gyr) owing to  the tidal torque of the halo. 
Note also that these spirals arms finally wind  to form a central gaseous ring
at $T$ = 10  ($\sim$ 2.7 Gyr). Results of models with
the lower inclination angle (0$^{\circ}$ $\le$  $\theta$ $\le$ 30$^{\circ}$)
are essentially the same as that of this model.
There is only one inner Lindblad resonance (ILR) point for all models but those  with
retrograde pattern speed. The points  of ILR ($R_{\rm ILR}$), corotation ($R_{\rm CR}$),
and outer Lindblad resonance ($R_{\rm OLR}$) are 
6.8, 14.2, and 20.7 kpc for this standard  model with ${\Omega}_{\rm P}$ 
= 3.84 km s$^{-1}$ kpc$^{-1}$.
Triaxial dark matter halos with 
figure rotation can be formed after major merging between
two NFW halos, if the remnant halos have larger spin parameters ($\lambda$) of $>$ $0.08$. 
}
\end{figure}

Figure 1 describes how gaseous spiral arms are formed, as the triaxial halo
rigidly rotates in the standard model (Model 1).
Owing to the difference in angular speed between the gas disk
and the triaxial halo, the gas disk continuously suffer from  the strong 
tidal force of the halo. 
As a natural result of this, two open trailing arms are 
gradually developed in the entire disk  within 1 Gyr ($T$ = 4). 
These two open arms quickly  wind with each other
to form a central high-density, ring-like
structure at $R$ $\sim$  0.3 (4.5 kpc) in our units ($T$ = 6 and  8).
Annular low-density gaseous regions 
(0.3 $\le$ $R$ $\le$ 0.5 in our units)  
forms just outside the inner ring
because of the disk's angular momentum redistribution caused by  the halo's torque
($T$ = 10).
Thus our simulations first confirms the early suggestion (Bureau et al. 1999) 
that the observed unusually extended spiral arms of NGC 2915
can be due to the tidal torque of the triaxial dark matter halo with figure rotation.

\section{Bar formation  by triaxial dark halos with figure rotation} 

Figure rotation of triaxial dark matter halos can significantly influence
not only the formation of stellar bars, spiral arms, and warps
but also gas dynamics and star formation histories in gas-rich galaxies
(Bekki \& Freeman 2002; Bekki \& Freeman 2003, in preparation).
Here we summarize briefly numerical results on stellar bar formation
(hereafter ``halo-triggered bar'' formation)
in disk galaxies embedded by triaxial (and spherically symmetric)
dark matter halos with and without figure
rotation (The details of the numerical models will be given in Bekki \& Freeman 2003). 
Based on fully self-consistent models (i.e., ``live'' halo,
and self-gravitating exponential disks),
we numerically investigated the shapes and the pattern speed
of bars formed in disks embedded by dark matter halos with
different pattern speeds of figure rotation, different radial density
profiles (e.g., with or without central cusps/cores), 
different mass ratios of halos to disks ($f_{\rm dm}$), and different
shapes of the halos. 

The results are summarized as follows:
(1) Stellar bars can be formed in disks embedded by
triaxial dark halos with figure rotation, even if $f_{\rm dm}$ 
is 5, for which no bars can form in models with
spherically symmetric halos,
(2) Triaxial shapes and figure rotation disappears
a few Gyr after the formation of stellar bars in the halos
owing to the mutual dynamical interaction (i.e., dynamical friction)
between disk and halo components,
(3) Stellar bars are less likely to form in dark halos with cuspy
radial distributions of the halos (i.e., bar formation
is more likely in halos with the density profiles
decribed by  Salucci \& Burkert 2000),
(4) The pattern speeds of the developed bars in
triaxial dark matter halos with figure rotation and with larger $f_{\rm dm}$ ($>$ 10)
are much lower than those  of bars formed 
in a spontaneous way  (i.e., due to bar instability)
for  massive disk models ($f_{\rm dm}  \sim 2$),
and (5) Morphological properties of halo-triggered bars
appear to be more similar  to those of ``tidal bars'' 
(Noguchi \& Ishibashi 1986) that can be formed in interacting/merging galaxies.

\section{Implications: Halo-triggered starbursts and bar formation at high redshifts} 

The above result (1) implies that bar formation is still possible
in low luminosity LSBs that are observed to be dominated by dark
matter halos, if the halos have figure rotation. 
If disk galaxies grow slowly by accretion of gas and satellite galaxies
over $\sim$ 10 Gyr,  
(younger) high-z disk galaxies 
are likely to have larger $f_{\rm dm}$
and thus be less likely to develop stellar bars within the disks
owing to spontaneous bar instability.
Therefore, if high-z disk galaxies have bars,  some of these disks
could be embedded by triaxial dark  halos with figure rotation.
The above result (2) implies that lower-z dark halos are less
likely to show triaxial shapes and figure rotation because
of the long-term dynamical interaction between the halo-triggered
bars and the halos. Regarding this point, 
BCD NGC 2915 and its triaxial halo could be formed relatively recently
(though we can not specify the possible physical reasons for its late
formation).

Non-axisymmetric barred structures (either in dark matter halos
or in stellar disks) have been suggested  to cause 
rapid and efficient radial gas transfer to the central
regions of galaxies and thus be responsible for
the formation of central  starbursts 
(e.g., Noguchi \& Ishibashi 1986; Bekki \& Freeman 2002).   
Our numerical results accordingly suggest that {\it (a) starbursts in galaxies can be
triggered by tidal force of their triaxial dark halos with figure rotation} and  
{\it (b) such halo-triggered starbursts are possible even in the  higher redshift universe
where disk galaxies are less strongly  self-gravitating and
spontaneous bar formation is thus highly unlikely, if the halos have figure rotation.}
We therefore suggest that star formation histories of disk galaxies can
be significantly different between those with and without triaxial
shapes and figure rotation of dark matter halos at high redshifts.
Lastly we stress that we have so far shown just two  examples of the possible
important roles of figure rotation of triaxial dark halos in galaxy evolution:
Several important gravitational effects  of these halos
are yet to be explored such as dynamical effects of these halos
on warp formation and on growth of central
bulges in disk galaxies.

\begin{figure}
\plotfiddle{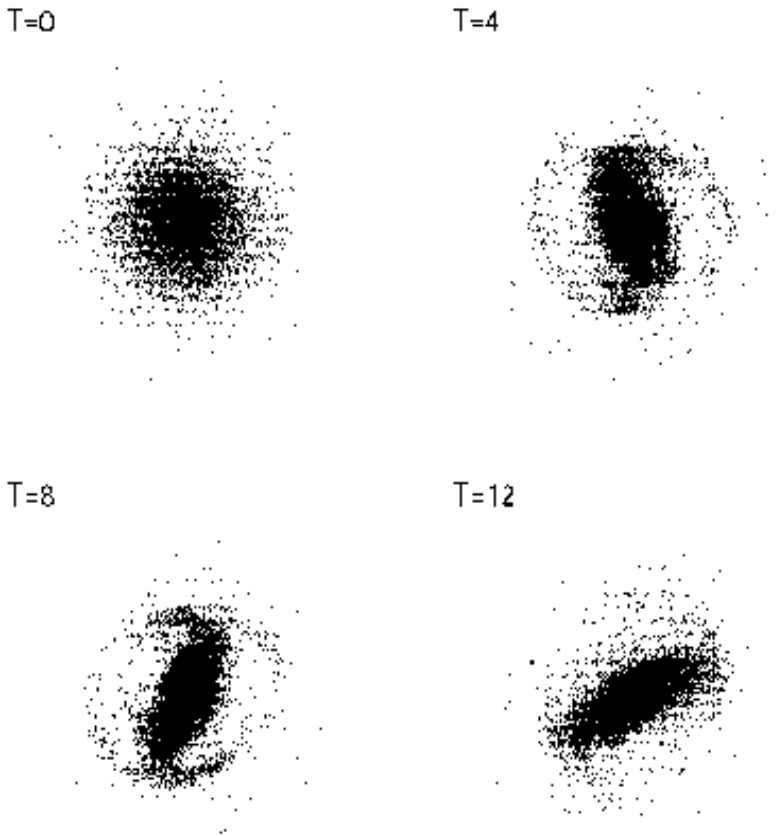}{5.in}{0}{80.}{85.}{-230.}{-50.}
\caption{
Morphological evolution of a stellar  disk projected onto the $x$-$y$ plane
for the disk galaxy model embedded by a massive dark matter halo
with a triaxial shape and figure rotation.
Triaxial dark halos with figure rotation can be constructed 
in the present numerical model 
based on numerical results on  major merging between two axisymmetric
spherical NFW (Navarro, Frenk \& White 1996) or SB (Salucci \&  Burkert 2000)
dark matter  halos {\it without figure rotation}. 
In this model, the spin parameter ($\lambda$) of the triaxial halo is $\sim$ 0.08,
which suggests that even if the spin parameter of a dark matter halo is not
so large, the halo can have figure rotation.
The time ($T$) indicated in the upper left
corner of each frame is given in our units (1.41 $\times$ $10^8$ yr) and
each frame (2.4 in our units) measures 42 kpc on a side.
Both the stellar  disk and the halo are assumed to rotate counter-clockwise
initially (The spin axis of the halo is coincident with the $z$ axis).
The mass ratio of the triaxial dark halo to the disk ($f_{\rm dm}$)
is 5 within 2 times  disk radius and
no bar can form in any spherically symmetric halo models with $f_{\rm dm}$ = 5.
Therefore, the stellar bar seen in this figure is due to the
tidal force of the massive triaxial halo with figure rotation for this model. 
The time scales of these halo-triggered bar formation in disks  depend on
$f_{\rm dm}$ and pattern speeds of figure rotation in the halos.
}
\end{figure}

\end{document}